%% file: Main.tex
\documentclass[twocolumn,10pt,twoside]{IEEEtran}

\usepackage{epsfig}
\usepackage{amsmath}
\usepackage{theorem}
\usepackage{amsmath}
\usepackage{mathrsfs}
\usepackage{amsfonts}
\usepackage{amssymb}
\usepackage{mathrsfs}
\usepackage{euscript}
\usepackage{graphicx}
\usepackage{multirow}
\usepackage{cite}
\usepackage{url}

\begin{document}

\title{Secrecy Capacity of Two-Hop Relay Assisted Wiretap Channels}
\author{
\authorblockN{Meysam Mirzaee, Soroush Akhlaghi\\}
\authorblockA{Shahed University, Tehran, Iran\\
Emails: {me.mirzaee,akhlaghi}@shahed.ac.ir
}
}
\maketitle

\begin{abstract}
Incorporating the physical layer characteristics to secure communications has received considerable attention in recent years. Moreover, cooperation with some nodes of network can give benefits of multiple-antenna systems, increasing the secrecy capacity of such channels. In this paper, we consider cooperative wiretap channel with the help of an Amplify and Forward (AF) relay to transmit confidential messages from source to legitimate receiver in the presence of an eavesdropper. In this regard, the secrecy capacity of AF relying is derived, assuming the relay is subject to a peak power constraint. To this end, an achievable secrecy rate for Gaussian input is evaluated through solving a non-convex optimization problem. Then, it is proved that any rates greater than this secrecy rate is not achievable. To do this, the capacity of a genie-aided channel as an upper bound for the secrecy capacity of the underlying channel is derived, showing this upper bound is equal to the computed achievable secrecy rate with Gaussian input. Accordingly, the corresponding secrecy capacity is compared to the Decode and Forward (DF) strategy which is served as the benchmark in the current work.
\end{abstract}


% For peer review papers, you can put extra information on the cover
% page as needed:
%\ifCLASSOPTIONpeerreview
%\begin{center} \bfseries EDICS Categories: COM-COOP, INF-CONF and INF-SECC \end{center}
%\fi
%
% For peerreview papers, this IEEEtran command inserts a page break and
% creates the second title. It will be ignored for other modes.
\IEEEpeerreviewmaketitle

\input{Introduction}
\input{System_Model.tex}
\input{Problem_Statement.tex}
\input{Simulation_Results.tex}
\input{Conclusion}
\input{App1.tex}
\input{App2.tex}
\input{App3.tex}

\input{App4.tex}
\input{App5.tex}

\bibliographystyle{IEEE}
% argument is your BibTeX string definitions and bibliography database(s)
\bibliography{Reference}
%\newpage

\end{document}

%% file: Introduction.tex
\section{Introduction}
\IEEEPARstart{S}{ecurity} has been regarded as one of the important issues in wireless communication networks as it may happen an illegitimate receiver to hear transmitted signal. As a result, enhancing security has attracted a great deal attentions in recent years in both of academia and industry. Information theoretic security is first proposed by Shannon in his landmark paper~\cite{Shannon} in which it is assumed both the legitimate receiver and eavesdropper (wiretapper) have direct access to the transmitted signal. Accordingly, using cryptographic approaches and the notion of equivocation, the level of uncertainty about the message and the key at the eavesdropper side is measured. However, this approach may not be feasible for some of wireless technologies~\cite{Schneier}. This motivated Wyner in his pioneering work in \cite{Wyner} to investigate the possibility of incorporating physical layer characteristics to secure the wireless communication networks.

Wyner introduced the wiretap channel in which a source wishes to send confidential message to a legitimate receiver while keeping the eavesdropper as ignorant of this information as possible when the broadcast channel between the source and the legitimate receiver and eavesdropper is a degraded one. Moreover, in~\cite{Wyner} the maximum achievable secrecy rate, the rate below which the message can not be decoded at the eavesdropper, is defined as the secrecy capacity. Accordingly, the secrecy capacity of discrete memoryless wiretap channels and Gaussian wiretap channels are investigated in \cite{Wyner} and \cite{Hellman}, respectively.

In \cite{Csiszár}, Csisz\'{a}r and K\"{o}rner generalized Wyner's approach to broadcast channels which are not necessarily degraded. It is assumed the source wishes to transmit a common message to both legitimate receiver and eavesdropper in addition to sending a confidential message to the legitimate receiver. This channel is termed as Broadcast Channel with Confidential message (BCC). Accordingly, both the capacity-equivocation region and the secrecy capacity region of BCC are established in \cite{Csiszár}.
Moreover, it is shown that in the lack of a common message,
the secrecy capacity can be computed as
\begin{equation}%\label{equ1}
    C_{s}=\max\; I(X;Y)-I(X;Z)
\end{equation}
where $X$, $Y$ and $Z$ are, respectively, the source input, the channel outputs at the legitimate receiver, and the eavesdropper's received signal where the maximization is taken over the distribution of channel input signal. Note that the secrecy capacity can be affected by channel conditions. For instance, if source-destination channel is weaker than source-eavesdropper channel the secrecy capacity will be zero meaning no confidential message can be transmitted. To overcome this issue, multiple antenna systems can be employed~\cite{Hero,Khisti1,Khisti2,Hassibi,Shafiee,Ekrem,Li}.

Due to the cost and size limitation, using multiple antennas at each node may not be practically feasible. Cooperative communications, however, is an effective way to get advantages of multi-antenna systems while incorporating single antenna nodes~\cite{Laneman,Sendonaris1,Sendonaris2,khajenouri,Shahan2,Shahbazpanahi1}. In cooperative communication, some nodes can act as intermediate nodes, dubbed relays, to facilitate the transmission between two nodes of network. Accordingly, there are some strategies to be employed at the relay nodes, among them, the Amplify and Forward (AF), and Decode and Forward (DF) are mostly addressed in the literature. In AF strategy, the relay sends an scaled version of its received signal to the destination without any more changes, while in DF, the relay attempts to decode the information, re-encodes again and transmits a coded version of information to the destination. As a result, the AF strategy is more simpler than DF. Furthermore, in some applications, the relay nodes may have low security level, thus it is desirable that transmitted messages to be confidential for the relays. These relays are called untrusted relays\cite{Oohama,He}. In such scenarios, the AF strategy is the prominent choice as the relay nodes do not need to have an access to the information bits, hence, they are unable to eavesdrop the information bits.

More recently, a great deal of attentions are devoted to the physical layer security issues in cooperative communication networks, where it is shown that relaying can improve the achievable secrecy rate of such networks\cite{Tekin,Gamal,Poor,Erkip}. For instance, in \cite{Dong} the secure communication for a source to destination with the help of multiple cooperating relays in the presence of one or more eavesdroppers is investigated by considering three cooperative strategies: (i) DF, (ii) AF and (iii) Cooperative Jamming (CJ). In \cite{Zhang}, the AF beamforming under total and individual relay power constraints is studied where the goal is maximizing the secrecy rate when perfect channel state information (CSI) is available. Moreover, the idea of relay selection for secure cooperative networks is considered in \cite{Chen,Ding,Krikidis}. Also, there are some related works on this issue~\cite{Dong,Li2,Mukherjee}.

In this paper, we derive the secrecy capacity of a simple cooperative wiretap channel in which a source wishes to send a confidential message to a legitimated destination with the help of an untrusted relay incorporating the AF strategy, where it is desirable to keep the information bits confidential from an eavesdropper. Referring to Fig.~\ref{fig:fig1}, it is assumed there is not direct link from source to destination and eavesdropper and the communication is occurred in two hops with the help of an AF relay in the middle of transmission. In this case, the received signal at the destination is a degraded version of the relay's. Thus, the DF strategy is optimal. However, we are interested in cases in which we are dealing with an untrusted relay which is unaware of incorporated codebook at the source and the AF strategy is employed at this node. In this regard, the secrecy capacity is fully characterized.

To this end, the achievable secrecy rate for Gaussian input is derived. Then, it is demonstrated that any rate greater than this rate is not achievable. Accordingly, the secrecy capacity is compared to the capacity of DF relaying to get an indication regarding the capacity loss due to the use of AF relaying.

The reminder of this paper is organized as follows. The system model is discussed in Section~\ref{sec:model}. Section~\ref{sec:problemstatement} provides the problem formulation followed by the main results, where some of technical details are provided in the Appendix. Numerical results are represented in Section~\ref{sec:sim}. Finally, Section~\ref{sec:conc} summarizes findings.

The following notations are used throughout this paper: We use bold upper and lower case characters for matrices and vectors, respectively. Symbols $h$ and $I$, respectively, denote differential entropy and mutual information. $\mathbb{R}^{n}$ is the set of all $n$-dimensional real-valued vectors. $\textbf{A}\succeq0$ means that $\textbf{A}$ is positive semi-definite matrix. Moreover, $E[x]$, $\text{Var}[x]$ and $\text{Cov}(x,y)$ denote the mean and variance of random variable $x$ and covariance of random variables $x$ and $y$, respectively. The notations $x^{*}$, $\Re\{x\}$ and $|x|$ refer to complex conjugate, real part and absolute value of complex variable $x$. Function $\{x\}^{+}$ is equivalent to $\max\{0,x\}$. Finally $x^{n}$ and $\mathcal{CN}\thicksim(0,\textbf{K})$ , respectively, denote a sequence of length $n$ and a zero-mean circularly symmetric complex Gaussian distribution with covariance $\textbf{K}$.

%% file: System_Model.tex
\section{System Model}\label{sec:model}
\begin{figure}
  \centering
  \includegraphics[width=8.00cm]{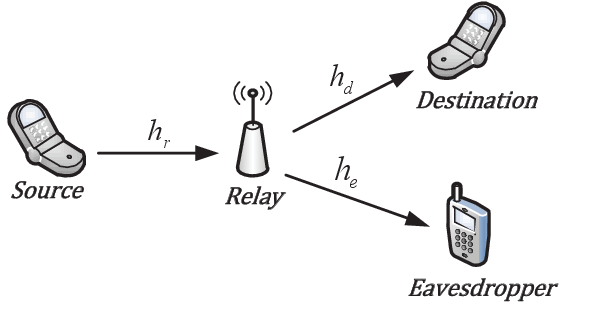}
  \caption{System Model}
  \label{fig:fig1}
\end{figure}

We consider a wireless communication network consisting of a source node S, a relay node R, a destination node D, and a passive eavesdropper E (see Fig. \ref{fig:fig1}). Moreover, it is assumed all nodes are equipped with single antenna and operate in half duplex mode. Also, it is assumed that there is not a direct link from S to D and E, and the communication is carried out in two hops through the use of a relay in the middle of transmission. We consider a quasi-static flat fading environment where all channel coefficients are assumed to be statistically independent. Moreover, in addition to the source-to-relay channel gain, the channel gains from the relay to the destination and eavesdropper are assumed to be completely known at the relay. This is in accordance to what is assumed in some of related works including~\cite{Bloch}.

According to the model depicted in Fig.~\ref{fig:fig1}, the communication is occurred in two hops. During the first hop, S sends the message \emph{W}, which is uniformly taken from the index set $\mathcal{W}=\{1,2,...,2^{nR}\}$, to the relay over a transmission interval of length $n$, where $R$ and $nR$ indicate, respectively, the transmission rate of source in units of bits per channel use and the message entropy. The mapping of each message $W$ to a codeword $x_{s}^{n}\in\chi_{s}^{n}$ is done by an encoder $f_{n}: \mathcal{W}\rightarrow \chi_{s}^{n}$, where $\chi_{s}^{n}$ is the transmitted vector space. Each source symbol $x_{s}(t)$, which appears within one time slot, has zero mean and unit power, i.e., $E\left[|x_{s}|^{2}\right]=1$. In this case, the received signal at the relay node can be written as,
\begin{equation}
y_{r}(t)=\sqrt{P_{s}}h_{r}x_{s}(t)+z_{r}(t)~~\textrm{for}~t=1,\ldots,n~,
\end{equation}
where $h_{r}$ is the channel fading coefficient from source to the relay, $z_{r}$ is zero-mean Additive White Gaussian Noise (AWGN) at the destination which is of unit power, i.e., $E\left[|z_{r}|^{2}\right]=1$. Finally, $P_{s}$ is the transmit power per symbol.

Then, the relay depending on the incorporated strategy, broadcasts a variation of the received information to the destination as well as the eavesdropper. In the following subsections, two relaying strategies, AF and DF, are investigated.

\subsection{Amplify and Forward}
In AF relaying, the relay transmits a scaled version of its received signal, i.e., $y_r$, to the destination as follows\footnote{For notational convenience, we ignore the index of symbols in the rest of paper.},
\begin{equation}\label{xr}
x_{r}=\omega y_{r}~,
\end{equation}
where $x_r$ is the transmitted signal and $\omega$ is a scaling factor, ensuring the peak power constraint at the relay is satisfied. As a result, the received signals at the nodes D and E can be written, respectively, as,
\begin{align}
y_{d}=h_{d}x_{r}+z_{d}=\sqrt{P_{s}}h_{d}\omega h_{r}x_{s}+h_{d}\omega z_{r}+z_{d}\label{yd}~,\\
y_{e}=h_{e}x_{r}+z_{e}=\sqrt{P_{s}}h_{e}\omega h_{r}x_{s}+h_{e}\omega z_{r}+z_{e}\label{ye}~,
\end{align}
where $h_{d}$ and $h_{e}$ are channel fading coefficients from R to D and E, respectively. Also, $z_{d}\sim \mathcal{CN}(0,1)$ and $z_{e}\sim \mathcal{CN}(0,1)$ are additive white Gaussian noises at the nodes D and E, respectively.

\subsection{Decode and Forward}
In DF strategy, the relay attempts to decode the source message and re-encodes the estimated message $W$ to a codeword $x_{r}^{n}\in\chi_{r}^{n}$ by an encoder $g_{n}: \mathcal{W}\rightarrow \chi_{r}^{n}$. For large transmission interval $n$, invoking the channel coding theorem, the relay can correctly decode the information signal as long as the transmission rate is not greater than the capacity of source-relay link, which is given by,
\begin{equation}\label{C_S_R}
C_{S-R}=\log_{2}\left(1+P_{s}|h_{r}|^2\right)~.
\end{equation}
After re-encoding, the relay broadcasts a weighted version of re-encoded symbols, i.e., $\omega x_{r}$, to D and E. Thus, the received signals at the nodes D and E can be respectively expressed as,
\begin{align}\label{yd_ye_DF}
y_{d}=h_{d}\omega x_{r}+z_{d}~,\\
y_{e}=h_{e}\omega x_{r}+z_{e}~.
\end{align}
In both AF and DF strategies, we assume that the relay is subject to a peak power constraint, i.e. $E[|x_{r}|^2]\leq P_{r}$ in AF and $E[|\omega x_{r}|^2]\leq P_{r}$ in DF. Thus, the scaling factor at the relay should satisfy the following constraints,
\begin{eqnarray}\label{power_constraint}
|\omega|^{2}\leq&\frac{P_r}{1+|h_{r}|^{2}P_{s}}\quad\text{for~ AF}~,\nonumber\\
|\omega|^{2}\leq& P_r~\quad\quad\quad\text{for~ DF}~,
\end{eqnarray}
where $E[|x_{r}|^2]=1$ is assumed in DF strategy.

In the sequel, we are going to compute the secrecy capacity of this network.

%% file: Problem_Statement.tex
\section{Secrecy capacity of channel}\label{sec:problemstatement}
This section aims to address the secrecy capacity of cooperative wiretap channel when the relay makes use of AF and DF strategies which are addressed in subsections III.A and III.B, respectively.

\subsection{Amplify and Forward}
The Amplify and Forward cooperative wiretap single-input single-output channel can be thought as a degraded broadcast channel. Hence, the secrecy capacity of this channel can be computed as~\cite{Elgamal},
\begin{align}\label{Cs_AF}
C_{s}(P_r)=\left\{\max_{\stackrel{E\{|x_{r}|^{2}\}\leq P_{r}}{p(x_{s})\in\rho}} \frac{1}{2}\Bigl[I(x_{s};y_{d})-I(x_{s};y_{e})\Bigr]\right\}^{+}~,
\end{align}
where $p(x_{s})$ is the Probability Density Function (PDF) of $x_{s}$ and $\rho$ is the set of all possible PDFs associated with zero mean/unit variance random variables. Also, the factor $\frac{1}{2}$ is due to the use of half-duplex nodes and the transmission is done during two time slots.

Evaluating the secrecy capacity of underlying channel using (\ref{Cs_AF}) may be computationally infeasible. This motivated us to propose the following theorem which aims at addressing this issue using an indirect approach.

\vspace{5pt}
\emph{\bf Theorem 1}: The secrecy capacity of cooperative amplify and forward wiretap channel is given by,
\begin{align}\label{Cs_final_AF}
    C_{s}&(P_{r})=\nonumber \\
    &\begin{cases}
        0&\alpha\leq \beta \\
            \frac{1}{2}\log_{2}\left(\frac{\alpha\beta P_{r}^{2}+(\alpha\mu+\beta)P_{r}+\mu}{\alpha\beta
                P_{r}^{2}+(\alpha+\beta\mu)P_{r}+\mu}\right)
            &\alpha>\beta \;\text{and}\; P_{r}\leq \sqrt{\frac{\mu}{\alpha\beta}} \\
            \frac{1}{2}\log_{2}\left(\frac{2\sqrt{\alpha\beta\mu}+\alpha\mu+\beta}{2\sqrt{\alpha\beta\mu}+\alpha+\beta\mu}\right)
            &\alpha>\beta \;\text{and}\; P_{r}>\sqrt{\frac{\mu}{\alpha\beta}}~,
          \end{cases}
\end{align}
where $\alpha=|h_{d}|^{2}, \beta=|h_{e}|^{2} \text{ and } \mu=1+P_{s}|h_{r}|^{2}$.

\emph{Proof}: We prove the above theorem in two steps. First, using (\ref{Cs_AF}), it is shown that (\ref{Cs_final_AF}) is achievable for Gaussian distribution. Next, for the converse part, we propose an upper bound and show that any transmission rate greater than (\ref{Cs_final_AF}) is not achievable.
%%%%%%%%%%%%%%%%%%%%%%%%%%%%%%%%%%%%%%%%%%%%%%%%%%%%%%    STEP 1    %%%%%%%%%%%%%%%%%%%%%%%%%%%%%%%%%%%%%%%%%%%%%%%%%%%%%%%%%%%%%
\subsubsection{The achievability of (\ref{Cs_final_AF})}
For Gaussian input, the achievable secrecy rate can be computed as,
\begin{align}\label{R_s.1}
    R_{s}(P_r)=\left\{\max_{E\{|x_{r}|^{2}\}\leq P_{r}}
    \frac{1}{2}\Bigl[I(x_{s};y_{d})-I(x_{s};y_{e})\Bigr]\right\}^{+}~.
\end{align}
Thus, referring to (\ref{yd}) and noting $x_s\sim\mathcal{N}(0,1)$, it follows,
\begin{align}\label{I_xs_yd}
    I(x_{s};y_{d})&=\text{log}_{2}\left(1+\frac{P_{s}|h_{d}|^{2}|\omega|^{2}|h_{r}|^{2}}{1+|h_{d}|^{2}|\omega|^{2}}\right)\nonumber\\
                  &=\text{log}_{2}\left(\frac{1+\alpha\mu|\omega|^{2}}{1+\alpha|\omega|^{2}}\right)~.
\end{align}
Similarly, noting (\ref{ye}), one can arrive at the following,
\begin{equation}\label{I_xs_ye}
    I(x_{s};y_{e})=\text{log}_{2}\left(\frac{1+\beta\mu|\omega|^{2}}{1+\beta|\omega|^{2}}\right)~.
\end{equation}
Substituting (\ref{I_xs_yd}) and (\ref{I_xs_ye}) into (\ref{R_s.1}), it turns out that the achievable secrecy rate becomes,
\begin{align}\label{R_s.2}
    R_{s}&(P_r)=\nonumber\\
      &\left\{\max_{|\omega|^{2}\leq \frac{P_{r}}{\mu}}\frac{1}{2}\text{log}_{2}
        \left(\frac{\alpha\beta\mu|\omega|^{4}+(\alpha\mu+\beta)|\omega|^{2}+1}{\alpha\beta\mu|\omega|^{4}+(\alpha+\beta\mu)|\omega|^{2}+1}\right)\right\}^{+}~.
\end{align}
To address the optimal solution of (\ref{R_s.2}), the following maximization problem should be tackled,
\begin{equation}\label{max_prob.1}
\max_{|\omega|^{2}\leq \frac{P_r}{\mu}}\frac{\alpha\beta\mu|\omega|^{4}+(\alpha\mu+\beta)|\omega|^{2}+1}{\alpha\beta\mu|\omega|^{4}+(\alpha+\beta\mu)|\omega|^{2}+1}~,
\end{equation}
which can be reformulated as,
\begin{align}\label{max_prob.2}
    \max_{x}f(x)&=\frac{\alpha\beta\mu x^{2}+(\alpha\mu+\beta)x+1}{\alpha\beta\mu x^{2}+(\alpha+\beta\mu)x+1}\nonumber~,    \\
    &\text{subject to  } 0\leq x \leq X
\end{align}
where $x=|\omega|^2$ and $X=\frac{P_{r}}{\mu}$. Although the objective function of (\ref{max_prob.2}) is the ratio of two convex quadratic functions, this function is not convex in general~\cite{Tuy,Gotoh}; hence, the method of Lagrange Multipliers does not give the optimal solution. To find the optimal value of $x$, i.e., $\hat{x}$, we consider two possible cases of $\alpha\leq \beta$ and $\alpha>\beta$ as the following.

\emph{Case} $\alpha\leq \beta$: In this case, we show that the optimal solution of (\ref{max_prob.2}) is $\hat{x}=0$. To this end, noting the definition of $\mu$, indicating $\mu\geq1$, it follows,
\begin{equation}
\alpha(\mu-1)\leq \beta(\mu-1)~,
\end{equation}
or equivalently,
\begin{equation}
\alpha\mu+\beta\leq \alpha+\beta\mu~.
\end{equation}
Thus, for $0 < x\leq X$ and noting $f(x)=\frac{\alpha\beta\mu x^{2}+(\alpha\mu+\beta)x+1}{\alpha\beta\mu x^{2}+(\alpha+\beta\mu)x+1}$, it turns out that the denominator of $f(x)$ is greater than the nominator. Therefore, we have $f(x)< 1$. On the other hand, since $f(0)=1$, the optimal value of $x$ becomes,
\begin{equation}\label{x*}
\hat{x}=0~.
\end{equation}

\emph{Case} $\alpha>\beta$: In this case, we show that the optimal value of (\ref{max_prob.2}) can be computed as,
\begin{equation}\label{x*.2}
    \hat{x}=
          \begin{cases}
            \frac{P_{r}}{\mu} &P_{r}\leq \sqrt{\frac{\mu}{\alpha\beta}} \\
            \frac{1}{\sqrt{\alpha\beta\mu}} &P_{r}>\sqrt{\frac{\mu}{\alpha\beta}}~,
          \end{cases}
\end{equation}
where $\hat{x}$ is derived through using the following theorem.

\vspace{5pt}
\emph{Theorem 2}: We consider the following optimization problem,
\begin{align}\label{max_sam}
    \max_{\;\textbf{x}\in \mathbb{R}^{n}}f(\textbf{x})&=\frac{\textbf{x}^{T}\textbf{Qx}+\textbf{q}^{T}\textbf{x}+q^{0}}
                                {\textbf{x}^{T}\textbf{Px}+\textbf{p}^{T}\textbf{x}+p^{0}}~,
\end{align}
where $\textbf{P}$ and $\textbf{Q}$ are $n\times n$ symmetric positive semi-definite matrices. To address the optimal solution, we define the following function,
\begin{equation}\label{F}
    F(\textbf{x},\lambda)=\textbf{x}^{T}\textbf{Qx}+\textbf{q}^{T}\textbf{x}+q^{0}
                             -\lambda(\textbf{x}^{T}\textbf{Px}+\textbf{p}^{T}\textbf{x}+p^{0}),~\lambda>0~.
\end{equation}
Also, we define the functions,
\begin{equation}\label{x(lambda).1}
    \textbf{x}(\lambda)=\text{arg} \max_{\;\textbf{x}\in \mathbb{R}^{n}} F(\textbf{x},\lambda)~\forall\lambda>0~,
\end{equation}
and
\begin{equation}\label{pi(lambda)}
    \pi(\lambda)=\max_{\;\textbf{x}\in \mathbb{R}^{n}} F(\textbf{x},\lambda)=F(\textbf{x}(\lambda),\lambda)~.
\end{equation}
If there exists $\hat{\lambda}>0$ for which $\pi (\hat{\lambda})=0$, then $\hat{\textbf{x}}\equiv \textbf{x}(\hat{\lambda})$ is the optimal solution of (\ref{max_sam}).

\emph{Proof}: see \cite{Gotoh}.

According to the Theorem 2 and referring to (\ref{max_prob.2}), we define $F(x,\lambda)$ as follows,
\begin{equation}\label{max_F}
F(x,\lambda)=\alpha\beta\mu(1-\lambda)x^{2}+\Bigl(\alpha\mu+\beta-\lambda(\alpha+\beta\mu)\Bigr)x+1-\lambda~,
\end{equation}
where we assume $\lambda>0$ and $0\leq x\leq X$.

\emph{Claim 1}: The optimal value of $\lambda$, i.e., $\hat{\lambda}$, falls in the interval $[1,\frac{\alpha\mu+\beta}{\alpha+\beta\mu})$.

\emph{Proof}: see appendix I.

Based on claim 1, it is sufficient to merely investigate $F(x,\lambda)$ for $1\leq \lambda<\frac{\alpha\mu+\beta}{\alpha+\beta\mu}$.

\begin{figure}
  \centering
  \includegraphics[width=8.00cm]{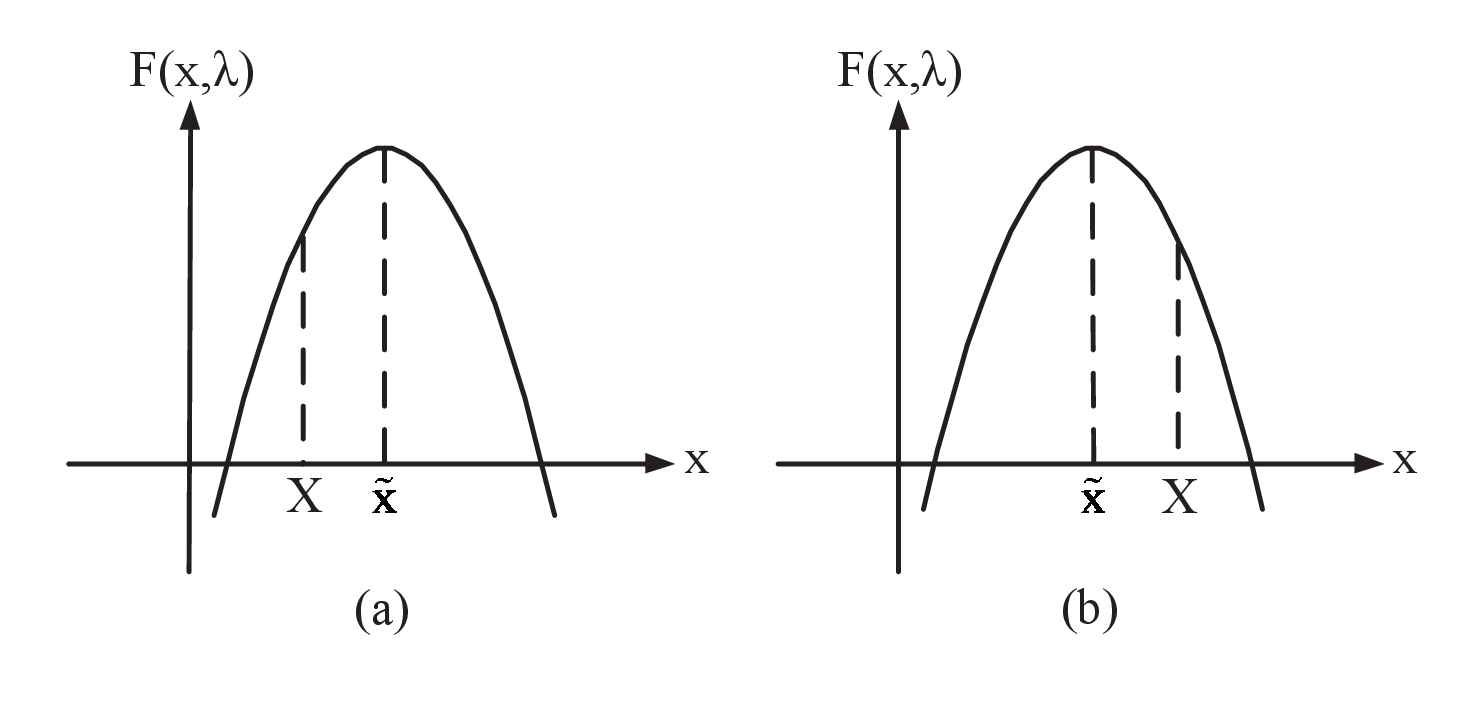}
  \caption{Illustration of function $F(x,\lambda)$ for two possible cases}
  \label{fig:fig2}
\end{figure}

It should be noted that referring to (\ref{max_F}), since $1-\lambda\leq 0$, it turns out that $F(x,\lambda)$ is a concave function of $x$ and has two positive roots\footnote{The number of positive roots of a polynomial with real coefficients ordered in terms of ascending power of the variable is either equal to the number of variations in sign of consecutive non-zero coefficients or less than this by a multiple of 2\cite{Anderson}}. As a result,  depending on the value of $\lambda$, $F(x,\lambda)$ can be represented as one of the curves illustrated in Fig.~\ref{fig:fig2}. Assuming $\tilde{x}$ maximizes $F(x,\lambda)$, if $X$ is equal or less than $\tilde{x}$, then $x(\lambda)=X$ gives the maximum value of $F(x,\lambda)$ in the interval $x\in[0,X]$~\big(see Fig.~\ref{fig:fig2}(a)\big); otherwise, $x(\lambda)$ will be equal to $\tilde{x}$~\big(see Fig.~\ref{fig:fig2}(b)\big).
Thus, we have,
\begin{equation}\label{x(lambda).2}
    x(\lambda)=
          \begin{cases}
            X &X\leq\tilde{x} \\
            \tilde{x} &X>\tilde{x}~,
          \end{cases}
\end{equation}
where $\tilde{x}$ is computed by taking derivation of $F(x,\lambda)$ with respect to $x$ and equating to zero as follows,
\begin{equation}\label{x_tilda}
\tilde{x}=\frac{\lambda(\alpha+\beta\mu)-(\alpha\mu+\beta)}{2\alpha\beta\mu(1-\lambda)}~,
\end{equation}
As a result, using~(\ref{x_tilda}) and claim 1, (\ref{x(lambda).2}) can be expressed as,
\begin{equation}\label{x(lambda).3}
    x(\lambda)=
          \begin{cases}
            X &1\leq\lambda\leq\frac{2\alpha\beta\mu X+\alpha\mu +\beta}{2\alpha\beta\mu X+\alpha+\beta\mu} \\
            \frac{\lambda(\alpha+\beta\mu)-(\alpha\mu+\beta)}{2\alpha\beta\mu(1-\lambda)}
                    &\frac{2\alpha\beta\mu X+\alpha\mu +\beta}{2\alpha\beta\mu X+\alpha+\beta\mu}<\lambda <\frac{\alpha\mu+\beta}{\alpha+\beta\mu}~.
          \end{cases}
\end{equation}
Also, using (\ref{x(lambda).1}) and (\ref{pi(lambda)}), $\pi(\lambda)$ can be obtained by,
\begin{align}\label{pi(lambda).2}
    \pi(\lambda)&=F\Bigl(x(\lambda),\lambda\Bigr)\nonumber\\
        &=
        \begin{cases}
            \pi_{1}(\lambda) &1\leq\lambda\leq\frac{2\alpha\beta\mu X+\alpha\mu +\beta}{2\alpha\beta\mu X+\alpha+\beta\mu} \\
            \pi_{2}(\lambda) &\frac{2\alpha\beta\mu X+\alpha\mu +\beta}{2\alpha\beta\mu X+\alpha+\beta\mu}< \lambda <\frac{\alpha\mu+\beta}{\alpha+\beta\mu}~,
        \end{cases}
\end{align}
where
\begin{align}\label{pi1}
    \pi_{1}(\lambda)=\Bigl(-\alpha\beta\mu X^{2}-&\bigl(\alpha+\beta\mu\bigr)X-1\Bigr)\lambda\nonumber\\
            &+\alpha\beta\mu X^{2}+\bigl(\alpha\mu+\beta\bigr)X+1~,
\end{align}
and
\begin{equation}\label{pi2}
    \pi_{2}(\lambda)=\frac{\Bigl(\lambda(\alpha+\beta\mu)-(\alpha\mu+\beta)\Bigr)^{2}}{4\alpha\beta\mu(\lambda-1)}-\lambda+1~.
\end{equation}
Claim 2: $\hat{\lambda}$ can be written as,
\begin{equation}\label{lambda_*}
    \hat{\lambda}=
        \begin{cases}
            \lambda_{1}=\frac{\alpha\beta\mu X^{2}+(\alpha\mu+\beta)X+1}{\alpha\beta\mu X^{2}+(\alpha+\beta\mu)X+1}
                    &X\leq\frac{1}{\sqrt{\alpha\beta\mu}}\\
            \lambda_{2}=\frac{2(\alpha+\beta\mu)(\alpha\mu+\beta)-8\alpha\beta\mu-\sqrt{\Delta}}{2(\alpha-\beta\mu)^{2}}
                    &X> \frac{1}{\sqrt{\alpha\beta\mu}}~,
        \end{cases}
\end{equation}
where
\begin{equation}\label{delta}
    \Delta=(8\alpha\beta\mu-2(\alpha+\beta\mu)(\alpha\mu+\beta))^{2}-4(\alpha-\beta\mu)^{2}(\alpha\mu-\beta)^{2}~.
\end{equation}

Proof: see appendix II.

Finally, substituting (\ref{lambda_*}) into (\ref{x(lambda).3}) yields,
\begin{equation}\label{x_*}
    \hat{x}=
        \begin{cases}
            X&X\leq\frac{1}{\sqrt{\alpha\beta\mu}}\\
            \frac{\lambda_{2}(\alpha+\beta\mu)-(\alpha\mu+\beta)}{2\alpha\beta\mu(1-\lambda)}
                &X>\frac{1}{\sqrt{\alpha\beta\mu}}~.
        \end{cases}
\end{equation}
Moreover, comparing (\ref{x_*}) with (\ref{x(lambda).2}), one can arrive at the following,
\begin{equation}\label{x*_final}
    \hat{x}=
        \begin{cases}
            X&X\leq \frac{1}{\sqrt{\alpha\beta\mu}}\\
            \frac{1}{\sqrt{\alpha\beta\mu}}&X> \frac{1}{\sqrt{\alpha\beta\mu}}~,
        \end{cases}
\end{equation}
or equivalently, we have,
\begin{equation}\label{x*_final.2}
    \hat{x}=
        \begin{cases}
            \frac{P_{r}}{\mu}&P_{r}\leq \sqrt{\frac{\mu}{\alpha\beta}}\\
            \frac{1}{\sqrt{\alpha\beta\mu}}&P_{r}> \sqrt{\frac{\mu}{\alpha\beta}}~.
        \end{cases}
\end{equation}
As a result, noting $\hat{x}=|\omega_{opt}|^2$, it turns out that if $P_{r}> \sqrt{\frac{\mu}{\alpha\beta}}$, the relay doesn't use all of its available power. This is due to the fact that the relay sends a noisy version of $x_{s}$ and additional relay's transmit power may enhance the additive noise, thereby decreasing the secrecy rate.

Finally, using (\ref{x*}) and (\ref{x*_final.2}) and after some mathematics, one can readily observe that (\ref{Cs_final_AF}) is the achievable secrecy rate of AF relying for Gaussian input. In what follows, we are going to show that any rate greater than (\ref{Cs_final_AF}) is not achievable (the converse part), thereby (\ref{Cs_final_AF}) is actually the secrecy capacity of AF relaying.
%%%%%%%%%%%%%%%%%%%%%%%%%%%%%%%%%%%%%%%%%%%%%%%%%%   STEP 2: Converse   %%%%%%%%%%%%%%%%%%%%%%%%%%%%%%%%%%%%%%%%%%%%%%%%%%%%%%%%%
\subsubsection{The Converse approach}
For the converse part, we show that any rate greater than $R_{s}(P_{r})$ defined in (\ref{R_s.2}) is not achievable. To do this, we investigate the capacity of genie-aided channel as an upper bound on the secrecy capacity of underlying channel. Then, we show that this upper bound is tight for Gaussian distribution. The following lemma establishes the capacity of corresponding genie-aided channel, \\
\emph{Lemma 1~\cite{Khisti1}}: An upper bound on the secrecy capacity of cooperative wiretap channel is,
\begin{equation}\label{upper_bound.1}
C_{s}(P_r)\leq\max_{\stackrel{p(x_{s})\in\rho}{|\omega|^2\leq\frac{P_r}{\mu}}} \frac{1}{2}I(x_{s};y_{d}|y_{e})~.
\end{equation}
In what follows, we show that for AF relaying, the Gaussian distribution maximizes $I(x_{s};y_{d}|y_{e})$. To this end, we have,
\begin{equation}\label{I(xs,yd)}
    I(x_{s};y_{d}|y_{e})=h(y_{d}|y_{e})-h(y_{d}|x_{s},y_{e})~.
\end{equation}
The second term in the right hand side of \eqref{I(xs,yd)} can be expressed, using ($\ref{yd}$) and ($\ref{ye}$), as,
\begin{align}\label{h(y_dx_s,y_e)}
    h(y_{d}|x_{s},y_{e})&=h(\sqrt{P_{s}}h_{d}\omega h_{r}x_{s}+h_{d}\omega z_{r}+z_{d}|x_{s},y_{e})\nonumber\\
                        &=h(h_{d}\omega z_{r}+z_{d}|x_{s},y_{e})\nonumber\\
                        &=h(h_{d}\omega z_{r}+z_{d}|h_{e}\omega z_{r}+z_{e})~.
\end{align}
One can readily observe that (\ref{h(y_dx_s,y_e)}) does not depend on the distribution of $x_{s}$, thus, $p(x_{s})$ should be chosen
such that the first term in the right hand side of (\ref{I(xs,yd)}), i.e., $h(y_{d}|y_{e})$, is maximized. On the other hand, we have,
\begin{align}\label{h(yd|ye).4}
    h(y_{d}|y_{e})&\stackrel{a}{=}h(y_{d}-\alpha_{\texttt{LMMSE}}y_{e}|y_{e})\nonumber\\
                  &\stackrel{b}{\leq} h(y_{d}-\alpha_{\texttt{LMMSE}}y_{e})\nonumber\\
                  &\leq \log_2(\pi \text{e}\lambda_{\texttt{LMMSE}})~,
\end{align}
where (a) comes from the fact that adding a known value to a random variable doesn't change the entropy and (b) holds since we always have $h(y|x)\leq h(x)$. $\alpha_{\texttt{LMMSE}}$ is the corresponding coefficient of Linear Minimum Mean Square Error (LMMSE) estimation of $y_d$ by $y_e$ and $\lambda_{\texttt{LMMSE}}$ is the error variance conditional on knowing $y_e$, i.e., $E[|y_{d}-\alpha_{\texttt{LMMSE}}y_{e}|^2|y_e]$. The last inequality in (\ref{h(yd|ye).4}) is due to the fact that the maximum differential entropy is achieved by Gaussian distribution.

In the case that $y_{d}$ and $y_{e}$ are jointly Gaussian, the estimation error, i.e., $y_{d}-\alpha_{\texttt{LMMSE}}y_{e}$ is independent of every linear function of $y_{e}$~\cite{kailath}, thus for Gaussian input we have $h(y_{d}-\alpha_{\texttt{LMMSE}}y_{e}|y_{e})=h(y_{d}-\alpha_{\texttt{LMMSE}}y_{e})$. Noting, the maximum differential entropy is achieved by Gaussian distribution, hence, the inequalities in (\ref{h(yd|ye).4}) are held with equality for Gaussian input $x_s$ and $I(x_{s};y_{d}|y_{e})$ is maximized. So, we can rewrite (\ref{upper_bound.1}) as,
\begin{align}\label{upper_bound.2}
    C_{s}(P_r)&\leq\max_{\;|\omega|^2\leq\frac{P_r}{\mu}} \frac{1}{2}I(x_{s};y_{d}|y_{e})\nonumber\\
     &=\max_{\;|\omega|^2\leq\frac{P_r}{\mu}}\frac{1}{2}\log_2(\pi \text{e}\lambda_{\texttt{LMMSE}})\nonumber\\
     &\quad-\frac{1}{2}h(h_{d}\omega z_{r}+z_{d}|h_{e}\omega z_{r}+z_{e})~,
\end{align}
where $\lambda_{\texttt{LMMSE}}$ can be computed as \cite{kailath},
\begin{align}\label{lambda}
    \lambda_{\texttt{LMMSE}}&=\text{Var}(y_{d}-\alpha_{\texttt{LMMSE}}y_{e}|y_{e})\nonumber\\
                            &=\text{Var}(y_{d})-\frac{|\text{Cov}(y_{d},y_{e})|^{2}}{\text{Var}(y_{e})}~.
\end{align}
We assume that the received noises at nodes D and E, i.e., $z_{d}$ and $z_{e}$, are jointly Gaussian with covariance matrix $K_{\phi}$, i.e.,
\begin{equation}\label{K_phi}
    \begin{bmatrix}
      z_{d} \\
      z_{e} \\
    \end{bmatrix}\sim \mathcal{CN}(\underline{0},K_{\phi}),\quad
    K_{\phi}=\begin{bmatrix}
                  1 & \phi^{*} \\
                  \phi & 1
             \end{bmatrix}.
\end{equation}
Using (\ref{K_phi}) and after some mathematics, it turns out that (\ref{lambda}) can be computed as,
\begin{equation}\label{lambda_final}
    \lambda_{\texttt{LMMSE}}=\frac{1+(\alpha+\beta)\mu x-|\phi|^{2}-2\Re\{\mu xh_{d}h_{e}^{*}\phi\}}{1+\beta\mu x}~.
\end{equation}
The proof is provided in Appendix III.

Moreover, the second term in (\ref{upper_bound.2}) can be computed as,
\begin{align}\label{2nd_term.1}
    h(h_{d}&\omega z_{r}+z_{d}|h_{e}\omega z_{r}+z_{e})\nonumber\\
            &=h(h_{d}\omega z_{r}+z_{d},h_{e}\omega z_{r}+z_{e})-h(h_{e}\omega z_{r}+z_{e})\nonumber\\
            &=\log_{2}\frac{\pi \text{e}\left(1+(\alpha+\beta)x-|\phi|^{2}-2\Re\{xh_{d}h_{e}^{*}\phi\}\right)}{1+\beta x}~.
\end{align}
The proof is given in appendix IV.

Plugging (\ref{lambda_final}) and (\ref{2nd_term.1}) into (\ref{upper_bound.2}) and after some manipulations, it follows,
\begin{align}\label{upper_bound.3}
        C_{s}(P_r) \leq\nonumber\\
         \max_{\;0\leq x \leq X} &\frac{1}{2}\log_{2}\Biggl\{\dfrac{1+\beta x}{1+\beta\mu x}\nonumber\\
                                            &\times\frac{1+(\alpha+\beta)\mu x-|\phi|^{2}-2\Re\{\mu xh_{d}h_{e}^{*}\phi\}}
                                            {1+(\alpha+\beta)x-|\phi|^{2}-2\Re\{xh_{d}h_{e}^{*}\phi\}}\Biggr\}~.
\end{align}
Note that the covariance matrix $K_{\phi}$ should be positive semi-definite, i.e., $K_{\phi}\succeq0$. This results in,
\begin{equation}\label{phi_condition}
    |\phi|\leq 1~.
\end{equation}
Thus, (\ref{upper_bound.3}) yields an upper bound just for values of $\phi$ which satisfy (\ref{phi_condition}).

Proceeding, we again consider two cases $\alpha\leq\beta$ and $\alpha>\beta$. For each case, an upper bound of secrecy capacity is computed with a special value of $\phi$.

\emph{Case $\alpha\leq\beta$}:
In this case, we choose,
\begin{equation}\label{phi.1}
    \phi=\frac{h_{d}^{*}}{h_{e}^{*}}~.
\end{equation}
Noting,
\begin{equation}%\label{}
    |\phi|^{2}=\frac{\alpha}{\beta}\leq 1~,
\end{equation}
thus substituting (\ref{phi.1}) into (\ref{upper_bound.3}) gives the following upper bound,
\begin{align}%\label{}
    &C_{s}(P_r) \leq \nonumber \\
    &\max_{\;0\leq x \leq X} \frac{1}{2}\log_{2} \frac{(1+\beta x)(1-\frac{\alpha}{\beta}+(\beta-\alpha)\mu x)}{(1+\beta\mu x)(1-\frac{\alpha}{\beta}+(\beta-\alpha)x)} \nonumber \\
    &= \max_{\;0\leq x \leq X} \frac{1}{2}\log_{2} \frac{\beta\mu(\beta-\alpha)x^{2}+(\beta-\alpha)(\mu+1)x+1-\frac{\alpha}{\beta}}{\beta\mu(\beta-\alpha)x^{2}+(\beta-\alpha)(\mu+1)x+1-\frac{\alpha}{\beta}}\nonumber \\
    &=0~.
\end{align}
This results in,
\begin{equation}\label{Cs_final_AF.2}
    C_{s}(P_r)=0.
\end{equation}
\emph{Case $\alpha>\beta$}:
In this case, $\phi$ is set to,
\begin{equation}%\label{}
    \phi=\frac{h_{e}}{h_{d}}~,
\end{equation}
where we should note the following,
\begin{equation}\label{phi.2}
    |\phi|^{2}=\frac{\beta}{\alpha}< 1~.
\end{equation}
Substituting (\ref{phi.2}) into (\ref{upper_bound.3}), we arrive at the following,
\begin{align}\label{upper_bound.41}%\label{}
    C_{s}(P_r) &\leq \max_{\;0\leq x \leq X} \frac{1}{2}\log_{2} \frac{(1+\beta
                x)\left(1-\frac{\beta}{\alpha}+(\alpha-\beta)\mu x\right)}{(1+\beta\mu
                x)\left(1-\frac{\beta}{\alpha}+(\alpha-\beta)x\right)}\nonumber\\
          &=\max_{\;0\leq x \leq X} \frac{1}{2}\log_{2} \left(\frac{1+\beta x}{1+\beta\mu x}\times\frac{1+\alpha\mu x}{1+\alpha x}\right)\\
          &=\max_{\;0\leq x \leq X} \frac{1}{2}\log_{2} \frac{\alpha\beta\mu x^{2}+(\alpha\mu+\beta)x+1}{\alpha\beta\mu x^{2}+(\alpha+\beta\mu)x+1} \label{upper_bound.4}~,
\end{align}
where (\ref{upper_bound.41}) is proved in Appendix V.

Referring to (\ref{R_s.2}), it turns out that (\ref{upper_bound.4}) is actually an achievable rate for the underlying channel. Thus, we have,
\begin{equation}\label{Cs_final_AF.3}
    C_{s}(P_r)=\max_{0\leq x \leq X} \frac{1}{2}\log_{2} \frac{\alpha\beta\mu x^{2}+(\alpha\mu+\beta)x+1}{\alpha\beta\mu x^{2}+(\alpha+\beta\mu)x+1}~.
\end{equation}
Considering the obtained results in (\ref{Cs_final_AF.2}) and (\ref{Cs_final_AF.3}), Theorem 1 is proved.
%%%%%%%%%%%%%%%%%%%%%%%%%%%%%%%%%%%%%%%%%%%%%   Decode and Forward %%%%%%%%%%%%%%%%%%%%%%%%%%%%%%%%%%%%%%%%%%%%%%%%%%%%%%%%%%
\subsection{Decode and Forward}
For DF relaying, using max-flow min-cut theorem, it turns out that the secrecy capacity can be computed as,
\begin{align}\label{Cs_DF}
    C_{s}(P_r)=\frac{1}{2}\min\left\{C_{S-R},C_{s_{R-D}}\right\}~,
\end{align}
where as mentioned earlier $C_{S-R}$ is the capacity of source-to-relay and $C_{s_{R-D}}$ is the secrecy capacity of the second hop operating at full power which is given by~\cite{Khisti1},
\begin{eqnarray}\label{C_s_RD}
C_{s_{R-D}}&=&\Biggl\{\max_{{}^{~|\omega|^2\leq P_{r}}_{~ p(x_{r})\in\rho}}\Bigl[I(x_{r};y_{d})- I(x_{r};y_{e})\Bigr]\Biggr\}^{+}\nonumber\\
&=&\left\{\log_{2}\left(\frac{1+\alpha P_{r}}{1+\beta P_{r}}\right)\right\}^{+}~.
\end{eqnarray}
If $C_{s_{R-D}}\leq C_{S-R}$, then we have $C_s(P_r)=\frac{1}{2}C_{s_{R-D}}$, otherwise, the minimum value of $C_{s_{R-D}}$ and $C_{S-R}$ is equal to $C_{S-R}$ and in this case, the relay does not need to use all of its available power, i.e.,~$P_{r}$. In other words, when the secrecy capacity associated with the second hop is greater than the available information at the relay, the relay can simply adjust its power so that not to waste any more power. In this case, we have $C_{s_{R-D}}=C_{S-R}$, where referring to (\ref{power_constraint}), one can arrive at the following~\footnote{It is worth mentioning that $\frac{1+\alpha |\omega|^2}{1+\beta |\omega|^2}$ is an increasing function with respect to $|\omega|$ for $\alpha>\beta$, thus decreasing $|\omega|$ reduces the secrecy rate of the second hop.}~,
\begin{equation}\label{C_s_DF2}
    C_{S-R}=\log_{2}\left(\frac{1+\alpha |\omega|^2}{1+\beta |\omega|^2}\right)~.
\end{equation}
By noting (\ref{C_S_R}) and the definition of $\mu$, we get,
\begin{equation}\label{C_s_DF2}
|\omega|^2=\frac{\mu-1}{\alpha-\beta\mu}~.
\end{equation}
As a result, the secrecy capacity of DF relaying is given by
\begin{equation}\label{Cs_final_DF}
    C_{s}(P_{r})=
        \begin{cases}
            0&\alpha\leq \beta\\
            \frac{1}{2}\log_{2}\left(\frac{1+\alpha P_{r}}{1+\beta P_{r}}\right)&\alpha>\beta \text{ and } \frac{1+\alpha P_r}{1+\beta P_r}\leq\mu\\
            \frac{1}{2}\log_{2}\mu&\alpha>\beta \text{ and } \frac{1+\alpha P_r}{1+\beta P_r}>\mu~,
        \end{cases}
\end{equation}
and the optimum relay's power can be written as,
\begin{equation}\label{omega_DF}
    |\omega_{opt}|^2=
        \begin{cases}
            0&\alpha\leq \beta\\
            P_r        &\alpha>\beta \text{ and } \frac{1+\alpha P_r}{1+\beta P_r}\leq\mu\\
            \frac{\mu-1}{\alpha-\beta\mu} &\alpha>\beta \text{ and } \frac{1+\alpha P_r}{1+\beta P_r}>\mu~.
        \end{cases}
\end{equation}

%% file: Simulation_Results.tex
\section{Simulation Results}\label{sec:sim}
This section aims to provide some numerical results to illustrate the secrecy capacity versus the power budget at the relay for cooperative wire-tap relay channel employing the AF and DF strategies. Throughout the simulations, the channel coefficients of source-relay ($h_{r}$), relay-destination ($h_{d}$) and relay-eavesdropper ($h_{e}$) are assumed to be Rayleigh distributed. Also, the received noises at the relay, the destination and the eavesdropper are assumed to be circularly symmetric complex Gaussian random variables with zero mean and unit variance. Moreover, the results are derived for different values of relay-destination channel strengths $\sigma^{2}_{h_{d}}=1, 2, 4$ and $8$, while it is assumed $\sigma^{2}_{h_{r}}=\sigma^{2}_{h_{e}}=1$ throughout the simulations. Also, source transmit power is set to $P_{s}=10$dBW~\footnote{Please note that here it is assumed the transmit SNR at the source is 10dB. Thus, noting the received noise at this node is of unit power, thus the transmit power at the source becomes 10dBW. }.

\begin{figure}
  \centering
  \includegraphics[width=8.00cm]{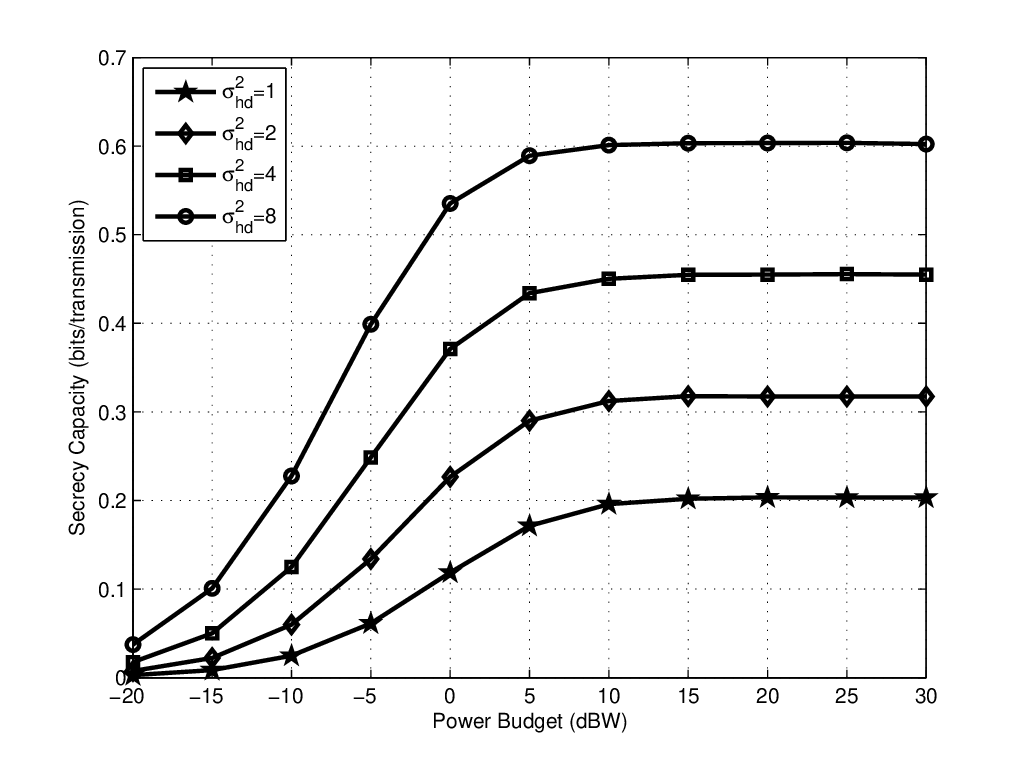}
  \caption{Secrecy capacity of AF relaying versus power budget}
  \label{fig:fig4}
\end{figure}

\begin{figure}
  \centering
  \includegraphics[width=8.00cm]{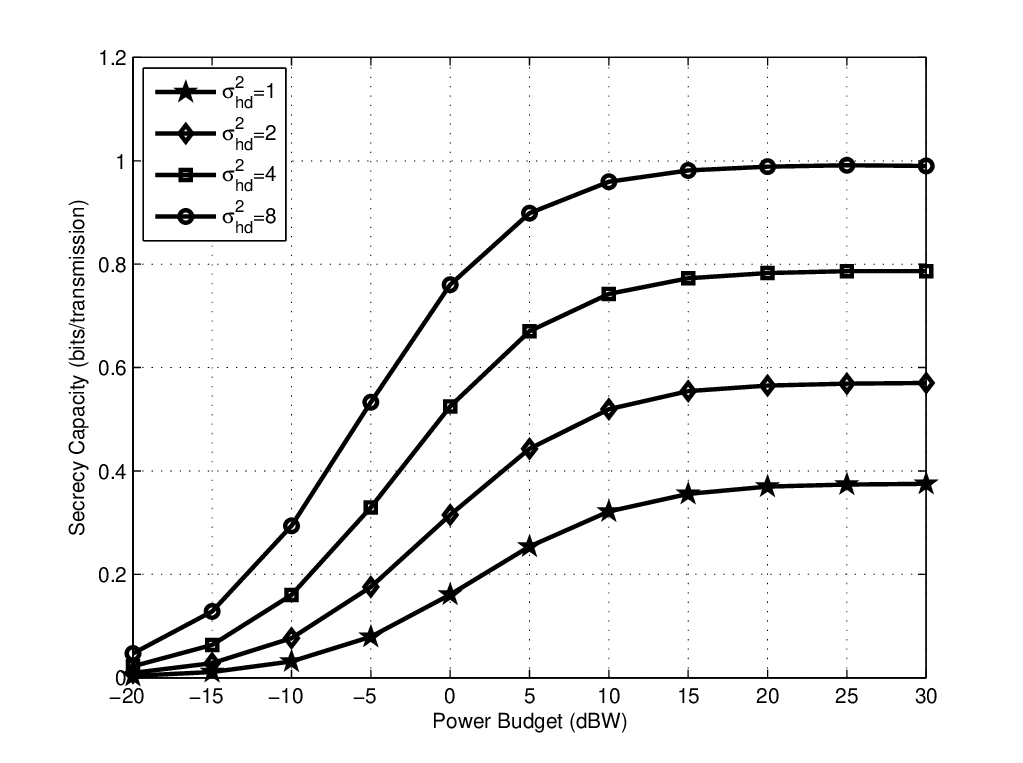}
  \caption{Secrecy capacity of DF relaying versus power budget}
  \label{fig:fig5}
\end{figure}

\begin{figure}
  \centering
  \includegraphics[width=8.00cm]{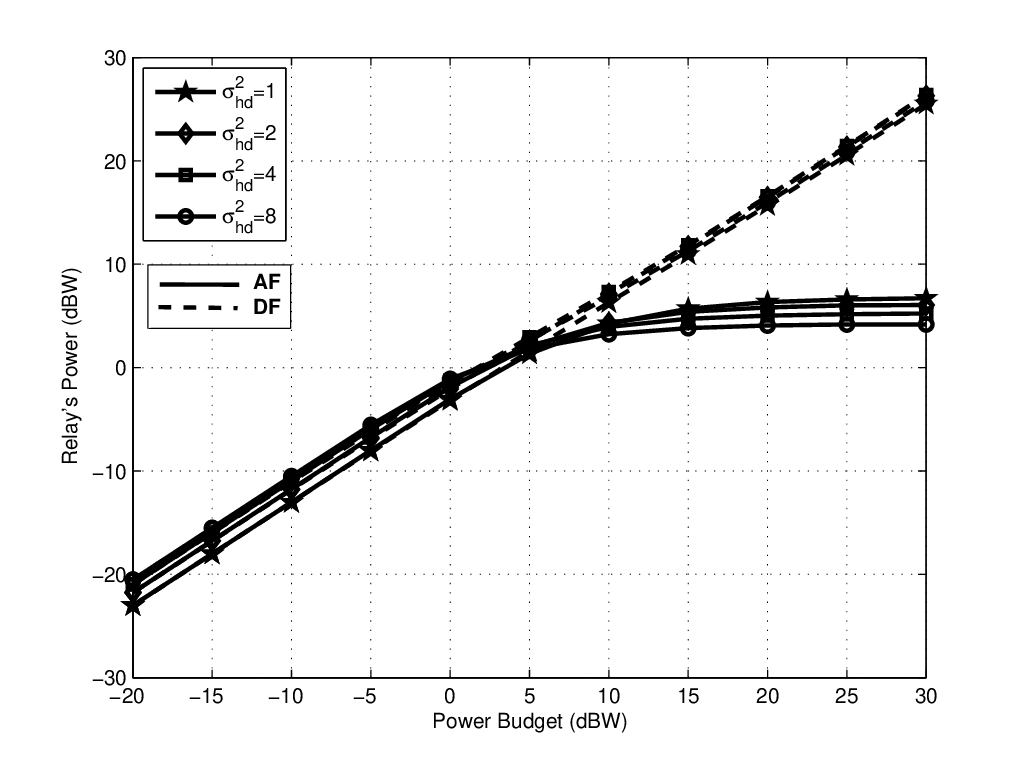}
  \caption{Relay's power for AF and DF strategies versus power budget}
  \label{fig:fig6}
\end{figure}

Figs. \ref{fig:fig4} and \ref{fig:fig5}, respectively, show the secrecy capacity of AF and DF cooperative wiretap channels versus power budget for various relay-destination channel strengths, implying the secrecy capacity of DF is greater than that of AF strategy. This is due to the fact that the received signal at the destination is a degraded version of the relay's, thus the DF strategy is optimal. Moreover, it is demonstrated that as the relay-destination channel strength is increased, the secrecy capacity is consistently increased. Moreover, the secrecy capacity approaches to a constant value as the relay's power tends to infinity. This is due to the fact that the capacity of the first hop acts as bottleneck. Also,~Fig.~\ref{fig:fig6} depicts the consumed relay's power versus available power for AF and DF strategies, showing the AF strategy saves more power as compared to DF strategy when the power budget at the relay increases.

%% file: Conclusion.tex
\section{Conclusion}\label{sec:conc}
This paper aimed at exploring the secrecy capacity of AF and DF relay-assisted wire-tap channel. Accordingly, the secrecy capacity of the aforementioned strategies are derived and numerically compared for Rayleigh channels. Although the secrecy capacity of DF relying outperforms that of AF, less power is consumed when relying on AF strategy.  

%% file: App1.tex
\section{Appendix I: Proof of claim 1}\label{sec:app1}
Note that $\hat{\lambda}$ should satisfy the equality $\pi(\hat{\lambda})=0$, where we have $f(\hat{x})=\hat{\lambda}$. Thus, the value of $\hat{\lambda}$ resides between lower bound and upper bound of $f(x)$. In the case of $\alpha > \beta$, we have,
\begin{equation}%\label{}
    \alpha\mu+\beta>\alpha+\beta\mu~,
\end{equation}
and therefore,
\begin{equation}%\label{}
    \alpha\beta\mu x^{2}+(\alpha\mu+\beta)x + 1 \geq \alpha\beta\mu x^{2}+(\alpha+\beta\mu)x+1~,
\end{equation}
where the equality is satisfied for $x=0$. Thus, we have $f(x)\geq1$.

On the other hand, we know that for positive values $a$, $b$ and $c$, when $b<a$, we have,
\begin{equation}%\label{}
\frac{a+c}{b+c}<\frac{a}{b}~.
\end{equation}
Based on this, by choosing $a=\alpha\mu+\beta$, $b=\alpha+\beta\mu$ and $c=\alpha\beta\mu x^{2}+1$, we arrive at,
\begin{equation}%\label{}
f(x)=\frac{\alpha\beta\mu x^{2}+(\alpha\mu+\beta)x + 1}{\alpha\beta\mu x^{2}+(\alpha+\beta\mu)x+1}<\frac{\alpha\mu+\beta}{\alpha+\beta\mu}~.
\end{equation}
Therefore, we conclude that,
\begin{equation}%\label{}
    1\leq \hat{\lambda}<\frac{\alpha\mu+\beta}{\alpha+\beta\mu}~.
\end{equation}

%% file: App2.tex
\section{Appendix II: proof of claim 2}\label{sec:app2}
\begin{figure}
  \centering
  \includegraphics[width=8.00cm]{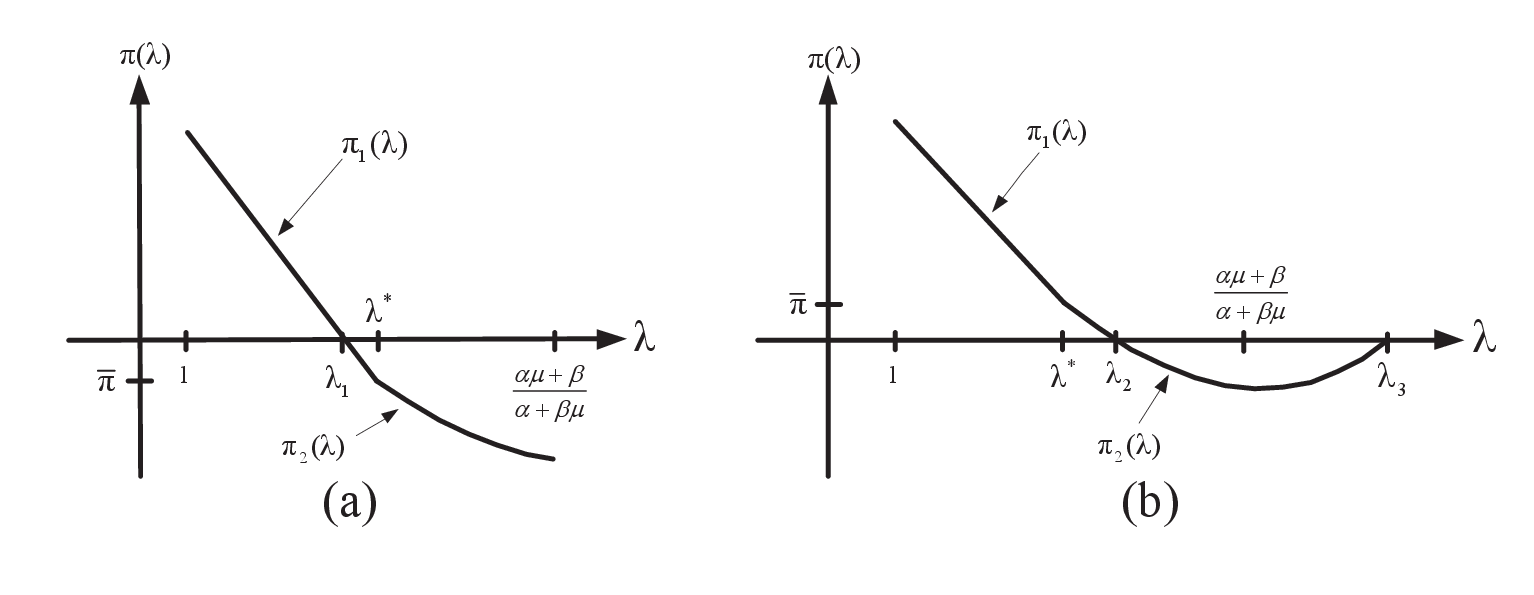}
  \caption{Illustration of function $\pi(\lambda)$ for two possible cases}
  \label{fig:fig3}
\end{figure}

Using (\ref{max_F}) and (\ref{x(lambda).3}), one can readily verify that $F(x,\lambda)$ and $x(\lambda)$ are continuous functions of $x$ and $\lambda$. Therefore, $\pi(\lambda)$ will also be a continuous function of $\lambda$. Furthermore, $\pi(\lambda)$ is a decreasing convex function of $\lambda$~\cite{Gotoh}. Moreover, $\pi(\lambda)$ has positive and negative values, respectively, at the start and end points of interval $\lambda\in[1,\frac{\alpha\mu+\beta}{\alpha+\beta\mu}]$, since from (\ref{pi1}) we have the following for $\lambda=1$,
\begin{equation}
\pi(1)=\pi_{1}(1)=(\alpha\mu+\beta)-(\alpha+\beta\mu)>0~,
\end{equation}
and for $\lambda=\frac{\alpha\mu+\beta}{\alpha+\beta\mu}$, using (\ref{pi2}), it follows,
\begin{equation}
\pi(\frac{\alpha\mu+\beta}{\alpha+\beta\mu})=\pi_{2}(\frac{\alpha\mu+\beta}{\alpha+\beta\mu})=1-\frac{\alpha\mu+\beta}{\alpha+\beta\mu}<0~.
\end{equation}
Thus, noting $\pi(\lambda)$ is strictly decreasing function, it has one root in the interval $\lambda\in[1,\frac{\alpha\mu+\beta}{\alpha+\beta\mu}]$, where this root should either reside in the region in which $\pi(\lambda)=\pi_{1}(\lambda)$ or $\pi(\lambda)=\pi_{2}(\lambda)$ as respectively illustrated in Fig.\ref{fig:fig3} (a) or Fig.\ref{fig:fig3} (b). To determine which of these conditions is occurred, we should compute $\pi(\lambda)$ at the point in which these curves meet each other, i.e., at the point $\lambda^*=\frac{2\alpha\beta\mu X+\alpha\mu+\beta}{2\alpha\beta\mu X+\alpha+\beta\mu}$ as illustrated in Fig.\ref{fig:fig3},
\begin{align}
    \bar{\pi}=\pi(\lambda^*) &= \pi_{1}(\frac{2\alpha\beta\mu X+\alpha\mu+\beta}{2\alpha\beta\mu X+\alpha+\beta\mu})\nonumber\\
    &=\frac{(\alpha\mu+\beta-(\alpha+\beta\mu))(\alpha\beta\mu X^{2}-1)}{2\alpha\beta\mu X+(\alpha+\beta\mu)}~.
\end{align}
This implies that for the case $X\leq\frac{1}{\sqrt{\alpha\beta\mu}}$, $\bar{\pi}$ has negative value, hence, $\pi(\lambda)$ can be represented as Fig.\ref{fig:fig3} (a). Thus, the root of $\pi(\lambda)$ can be computed through setting $\pi_1(\lambda)$ to zero as follows,
\begin{align}
    \pi_{1}(\lambda)=(-\alpha\beta\mu X^{2}&-(\alpha+\beta\mu)X-1)\lambda+\nonumber\\
                            &\alpha\beta\mu X^{2}+(\alpha\mu+\beta)X+1=0~,
\end{align}
which gives,
\begin{equation}
\lambda_{1}=\frac{\alpha\beta\mu X^{2}+(\alpha\mu+\beta)X+1}{\alpha\beta\mu X^{2}+(\alpha+\beta\mu)X+1}~.
\end{equation}
Alternatively, Fig.\ref{fig:fig3} (b) corresponds to the case that we have $X>\frac{1}{\sqrt{\alpha\beta\mu}}$. Consequently, the root of $\pi(\lambda)$ is derived by setting $\pi_{2}(\lambda)$ to zero as follows,
\begin{equation}%\label{}
    \pi_{2}(\lambda)=\frac{(\lambda(\alpha+\beta\mu)-(\alpha\mu+\beta))^{2}}{4\alpha\beta\mu(\lambda_2-1)}-\lambda+1=0~,
\end{equation}
or equivalently, we have,
\begin{equation}%\label{}
(\alpha-\beta\mu)^{2}\lambda^{2}+(8\alpha\beta\mu-2(\alpha+\beta\mu)(\alpha\mu+\beta))\lambda+(\alpha\mu-\beta)^{2}=0~,
\end{equation}
which has the following roots,
\begin{align}%\label{}
    \lambda_{2}=\frac{2(\alpha+\beta\mu)(\alpha\mu+\beta)-8\alpha\beta\mu-\sqrt{\Delta}}{2(\alpha-\beta\mu)^{2}}~,\\
    \lambda_{3}=\frac{2(\alpha+\beta\mu)(\alpha\mu+\beta)-8\alpha\beta\mu+\sqrt{\Delta}}{2(\alpha-\beta\mu)^{2}}~.
\end{align}
It is clear that $\lambda_{2}$ is the desirable root of $\pi_{2}(\lambda)$. Thus, we have,
\begin{equation}%\label{}
    \hat{\lambda}=
        \begin{cases}
            \frac{\alpha\beta\mu X^{2}+(\alpha\mu+\beta)X+1}{\alpha\beta\mu X^{2}+(\alpha+\beta\mu)X+1}
                &X\leq \frac{1}{\sqrt{\alpha\beta\mu}}\\
            \frac{2(\alpha+\beta\mu)(\alpha\mu+\beta)-8\alpha\beta\mu-\sqrt{\Delta}}{2(\alpha-\beta\mu)^{2}}
                &X>\frac{1}{\sqrt{\alpha\beta\mu}}~.
        \end{cases}
\end{equation}

%% file: App3.tex
\section{Appendix III}\label{sec:app3}
According to (\ref{yd}), we have,
\begin{align}\label{Var_yd}
    \text{Var}(y_{d})&=P_{s}|\omega|^{2}|h_{d}|^2|h_{r}|^2+|\omega|^2|h_{d}|^2+1\nonumber\\
              &=1+|\omega|^2|h_{d}|^2(1+P_{s}|h_{r}|^2)\nonumber\\
              &=1+\alpha\mu x~.
\end{align}
Similarly, using (\ref{ye}), it follows,
\begin{align}\label{Var_ye}
    \text{Var}(y_{e})=1+\beta\mu x~.
\end{align}
Also, the cross covariance of $y_{d}$ and $y_{e}$ can be computed as,
\begin{align}\label{cov_yd_ye}
    \text{Cov}(y_{d},y_{e})&=E\biggl[\left(\sqrt{P_{s}}h_{d}\omega h_{r}x_{s}+h_{d}\omega z_{r}+z_{d}\right)\nonumber\\
                    &\qquad\times\left(\sqrt{P_{s}}h_{e}\omega h_{r}x_{s}+h_{e}\omega z_{r}+z_{e}\right)^{*}\biggr]\nonumber\\
                    &=\left(1+P_{s}|h_{r}|^{2}\right)|\omega|^{2}h_{d}h_{e}^{*}+\phi^{*}\nonumber\\
                    &=\mu xh_{d}h_{e}^{*}+\phi^{*}~.
\end{align}
Using (\ref{Var_yd}), (\ref{Var_ye}) and (\ref{cov_yd_ye}), (\ref{lambda}) can be written as,
\begin{align}\label{lambda.2}
    \lambda_{\texttt{LMMSE}}&=1+\alpha\mu x-\frac{\alpha\beta\mu^{2}x^{2}+|\phi|^{2}+
                                        2\Re\{\mu xh_{d}h_{e}^{*}\phi\}}{1+\beta\mu x}\nonumber\\
                            &=\frac{1+(\alpha+\beta)\mu x-|\phi|^{2}-2\Re\{\mu xh_{d}h_{e}^{*}\phi\}}{1+\beta\mu x}~.
\end{align}

%% file: App4.tex
\section{Appendix IV}\label{sec:app4}
We begin with the following definition,
\begin{align}\label{Kn}
    K_{z}&\triangleq \text{Cov}\left(\begin{bmatrix}
                h_{d}\omega z_{r}+z_{d} \\
                h_{e}\omega z_{r}+z_{e} \\
              \end{bmatrix}
    \right)\nonumber\\
    &=\begin{bmatrix}
        |\omega|^{2}|h_{d}|^{2}+1 & |\omega|^{2}h_{d}h_{e}^{*}+\phi^{*} \\
        |\omega|^{2}h_{d}^{*}h_{e}+\phi & |\omega|^{2}|h_{e}|^{2}+1 \\
      \end{bmatrix}~.
\end{align}
We know that the following holds,
\begin{align}\label{2nd_term.2}
    h(h_{d}&\omega z_{r}+z_{d}|h_{e}\omega z_{r}+z_{e})=\log_{2}\frac{\pi \text{e}|K_{z}|}{\text{Var}(h_{e}\omega z_{r}+z_{e})}~,
\end{align}
where
\begin{align}\label{det}
    |K_{z}|=1+(\alpha+\beta)x-|\phi|^{2}-2\Re\left\{xh_{d}h_{e}^{*}\phi\right\}~,
\end{align}
and
\begin{align}
    \text{Var}\left(h_{e}\omega z_{r}+z_{e}\right)=\beta x+1~.
\end{align}
So, we arrive at (\ref{2nd_term.1}).

%% file: App5.tex
\section{Appendix V}\label{sec:app5}
To prove (\ref{upper_bound.41}), we use the following equality,
\begin{align}
    \frac{1+\alpha x}{1+\alpha\mu x}&\times
    \frac{1-\frac{\beta}{\alpha}+(\alpha-\beta)\mu x}{1-\frac{\beta}{\alpha}+(\alpha-\beta)x} \nonumber\\
& =\frac{\alpha\mu(\alpha-\beta)x^{2}+(\alpha-\beta)(1+\mu)x+1-\frac{\beta}{\alpha}}{\alpha\mu(\alpha-\beta)x^{2}+(\alpha-\beta)(1+\mu)x+1-\frac{\beta}{\alpha}}=1~.
\end{align}
Thus, we have,
\begin{align}
    \frac{1+\alpha\mu x}{1+\alpha x}=
    \frac{1-\frac{\beta}{\alpha}+(\alpha-\beta)\mu x}{1-\frac{\beta}{\alpha}+(\alpha-\beta)x}~,
\end{align}
which yields the equality of (\ref{upper_bound.41}).